\documentclass[aps,prd,nofootinbib,superscriptaddress,amsfonts,amssymb,amsmath,twocolumn,floatfix,natbiboptions=numbers]{revtex4-2}

\usepackage{soul}
\usepackage[normalem]{ulem} 
\usepackage{xcolor}         
\usepackage{amsfonts}       
\usepackage[makeroom]{cancel}
\usepackage{txfonts}
\usepackage{booktabs}

\usepackage{titlesec}
\usepackage{enumitem}
\usepackage[USenglish]{babel}
\usepackage{epsfig}
\usepackage{graphics}
\usepackage{graphicx}       
\usepackage{amsmath}
\usepackage{comment}
\usepackage{slashed}
\usepackage{siunitx}

\usepackage{hyperref}
\hypersetup{
    colorlinks=true,
    linkcolor=blue!50!black, 
    citecolor=blue!50!black,
    urlcolor=blue!50!black,
    filecolor=magenta,
    pdfpagemode=FullScreen,
}
\date{\today}

\begin{document}

\title{{Nonperturbative Danielson-Satishchandran-Wald Decoherence with Unruh-DeWitt detectors}}

\author{Levy B. N. Batista}
\email{L.DNBatista@swansea.ac.uk}
\affiliation{Centre for Quantum Fields and Gravity, Department of Physics, Swansea University, Singleton Park, Swansea, SA2 8PP, U.K.}

\author{Andr\'e G. S. Landulfo}
\email{andre.landulfo@ufabc.edu.br}
\affiliation{Centro de Ci\^encias Naturais e Humanas, Universidade Federal do ABC, Rua Santa Adélia, 166, 09210-170, Santo André, São Paulo, Brazil}

\author{Robert B. Mann}
\email{rbmann@uwaterloo.ca}
\affiliation{Department of Physics and Astronomy, University of Waterloo,
200 University Avenue West, Waterloo, Ontario N2L 3G1, Canada}

\author{George E. A. Matsas}
\email{george.matsas@unesp.br}
\affiliation{Instituto de Física Teórica, 
Universidade Estadual Paulista, Rua Dr. Bento Teobaldo Ferraz, 271, 01140-070, São Paulo, São Paulo, Brazil}


\begin{abstract}
  Recently, Danielson, Satishchandran, and Wald (DSW) have proposed a novel source of decoherence for uniformly accelerated charges and masses in spatial superposition in spacetimes containing a bifurcating Killing horizon. Such an effect can be traced back to the emission and absorption of soft photons and gravitons, which effectively act as ``which-path'' information probes. This results in the decoherence of any such superposition in a finite proper time. With this in mind, we study the DSW mechanism using a gapless finite-time detector prepared in a spatial superposition of uniformly accelerated paths in Minkowski spacetime that interacts with a scalar field. The calculation is nonperturbative. This model will enable us to analyze the decoherence process in a more controlled manner, highlighting the main factors that give rise to this interesting mechanism. 
\end{abstract}

\maketitle

     \section{Introduction}
     \label{introduction}


The decoherence of quantum superpositions in scenarios where gravity plays a significant role has recently attracted growing attention~(see, {\it e.g.}, Refs. \cite{ZCPB11,B13,AH13,PZCB15}). This is not only interesting in its own right, but also offers a promising avenue toward a deeper understanding of the quantum nature of gravity~\cite{AM25}, particularly in the midst of the debate over whether or not gravitons exist. Concerning this, Belenchia {\it et al.}, {\it e.g.}, have convincingly argued that the vacuum of spin-2 massless fields, associated with gravitons, must be quantized to maintain the consistency of quantum mechanics at low energies~\cite{BWGRBA18, DSW22a}.

The discussion above led Danielson, Satishchandran, and Wald~(DSW) to the interesting conclusion that charges and masses lying at rest in a quantum spatial superposition outside black holes would decohere due to their interaction with photons and gravitons, respectively, with negligible energy as measured by static observers at infinity~\cite{DSW22b,DSW23,DSW25}. This turns out to be a sort of memory effect imprinted at the Killing horizon, analogous to the more common memory effect of soft particles~\cite{SW65} impressed at null infinity due to scattering of electromagnetic and gravitational fields~\cite{B85,BT87,SZ16}. From a physical perspective, the DSW effect can be understood from the fact that the zero-energy photons and gravitons (concentrated at the Killing horizon) of the Unruh thermal bath that interact with the uniformly accelerated charges and masses, respectively, are eventually those responsible for the usual Larmor radiation as described by inertial observers~\cite{higuchi_1992R,higuchi_1992,portales_oliva_2022,vacalis_2023,portales_oliva_2024,brito_2024,UK_BR}. (Analogous results hold for black holes~\cite{HMS97,HMS98,CHM98}.) Thus, it should not come as a complete surprise that the soft photons and gravitons above function as ``which-path'' probes. 

Let us introduce the DSW mechanism, focusing on the electromagnetic case. Consider an experimenter, Alice, following the time isometry of a stationary spacetime, holding a charge with spin  $|\rightarrow \rangle$; as usual, $\sigma^x|\rightarrow \rangle=|\rightarrow \rangle$ and $\sigma^x$ is a Pauli matrix. The initial state of the electromagnetic field is a (quasifree) state $|\Omega \rangle$ invariant under the time isometry. At some moment, Alice uses a Stern-Gerlach process to create the superposition
\begin{equation}
    |\psi \rangle 
    \equiv 
    \frac{1}{\sqrt{2}} \Big(|\uparrow, \psi_1 \rangle  
    +
    |\downarrow,\psi_2 \rangle \Big)
    \label{particle's initial state}
\end{equation}
over a time interval $T_1$, where $\sigma^z|\uparrow \rangle=|\uparrow \rangle$, $\sigma^z|\downarrow \rangle=-|\downarrow \rangle$, and $|\psi_1 \rangle$ and $|\psi_2 \rangle$ are spatially separated normalized states. Alice holds the superposition for a time interval $T$ before recombining it over a time interval $T_2$, by using an inverse Stern-Gerlach process. At the end, charge and radiation  should evolve, in general, to the entangled state
\begin{equation}
    \frac{1}{\sqrt{2}} \Big(|\uparrow, \psi_1 \rangle \otimes | \phi_1 \rangle  
    +
    |\downarrow,\psi_2 \rangle \otimes |\phi_2 \rangle \Big),
    \label{initial state 2}
\end{equation}
where $|\phi_1 \rangle$ and $|\phi_2 \rangle$ are coherent states sourced by the charge. Therefore, the charge loses coherence as it becomes correlated with the radiation field. This can be quantified by verifying whether or not the final radiation states are distinguishable. It turns out that the decoherence is intimately related to the expected number of the so-called {\it entangling photons} emitted during the experiment. They act as a ``which-path" information probe, and their expected number depends on the choice of the vacuum state $|\Omega \rangle$.

Let us now consider the particular case in which Alice evolves freely in the inertial vacuum of Minkowski spacetime, carrying the superposition with her. In this case, photon emission occurs only during the forward and reverse Stern-Gerlach stages. This can be minimized by making the processes adiabatic. The picture changes, however, if, rather than evolving freely, Alice has proper acceleration $ a = {\rm const}$. Assuming that the spatial separation between the superposition components is $d\ll a^{-1}$ (as measured by Alice), the decoherence turns out proportional to the time $T$ (assuming $T \gg T_1, T_2$) over which the charge is uniformly accelerated and emits/absorbs soft Rindler photons (concentrated at the Killing horizon) to/from the Unruh thermal bath~\cite{higuchi_1992R,higuchi_1992,portales_oliva_2022,vacalis_2023}. The larger $a$ and $d$ are, the greater the decoherence. We should stress that a charge in spatial superposition, lying at rest with a usual (KMS) thermal state in Minkowski spacetime, does not decohere. (In contrast to the zero-energy Rindler modes, which can still be labeled by their {\it non-zero} transverse momentum ${\bf k}_\perp$, zero-energy Minkowski modes have ${\bf k}_\perp=0$.) 

The situation is similar if one considers the experiment in Schwarzschild spacetime with Alice's lab held static in the Hartle-Hawking or Unruh vacua. Again, the charge emits {/absorbs soft} photons ({concentrated at} the event horizon) {to/from Hawking radiation}~\cite{HMS97,HMS98,CHM98},  decohering the superposition as a consequence~\cite{DSW25}. On the other hand, as expected, no decoherence occurs in the (no-particle) Boulware vacuum, where the stimulated (as well as spontaneous) emission of zero-energy photons vanishes. 

To better understand the DSW decoherence mechanism, we replace the superposed charge coupled to the photon field with a superposed two-level system (also referred to as an Unruh-DeWitt detector) coupled to a scalar field. Superposed detectors have been used in a variety of situations \cite{Foo:2020dzt}, including for new tests of the Unruh effect \cite{Martin-Martinez:2010gnz}, probing causal ordering \cite{Henderson:2020zax}, elucidating observational imprints of quantum gravitational effects~\cite{Foo:2021exb}, and in deriving the so-called ``worldline-induced transparency''~\cite{Azizi2026}.  
 
Here, the forward and reverse Stern-Gerlach processes used to split and recombine the charge state are replaced by processes of switching the detector on and off. The switching must be (at least) continuous to avoid ultraviolet divergences~\cite{HMP93}. Moreover, we consider a {\it massive} scalar field to avoid some infrared divergences that arise when Alice is inertial. Finally, we degenerate the two-level gap, $\Delta E\to 0$, to constrain stationary detectors to interact only with zero-energy particles (as defined by comoving observers), as these are the source of decoherence in the DSW mechanism. As a great bonus, taking $\Delta E\to 0$ allows for a nonperturbative treatment of the system’s evolution~\cite{AL16}. 

We focus on the cases where the detector is (a1)~free in the (usual) inertial vacuum of Minkowski spacetime and (a2)~static in the Unruh thermal bath of the Rindler wedge, and compare the results with the DSW ones, where the charge is (b1)~free in the Minkowski vacuum and (b2)~static in the thermal bath of the Hartle-Hawking vacuum of the Schwarzschild black hole, respectively. We show that in case (a1) the decoherence arises solely due to particle emission along the switching process, which occurs at the cost of some external work~\cite{BL21}, which is analogous to the conclusion that in case~(b1) the decoherence is all due to photon emission along the forward and reverse Stern-Gerlach processes. On the other hand, in cases~(a2) and~(b2), the decoherence is mainly caused by soft particles emitted/absorbed to/from the corresponding horizons over time $T$ during which the systems stay in superposition. In addition, we compare our model with DSW's across various key parameters, including the duration for which the detector is active, proper acceleration, and separation between the superposition paths. We stress that our results are exact in the sense that no approximations are employed to extract leading-order contributions.

Our paper is organized as follows. In Sec.~\ref{sec:Model}, we present the setup, compute the exact time evolution operator using a Magnus expansion,  quantify the decoherence, and relate it to the number of entangling particles emitted/absorbed by the detector. In Secs.~\ref{sec:WaldRec0} and~\ref{sec:WaldRec+}, we show how the decoherence depends on the physical properties of the detector superposition, assuming that Alice is inertial and uniformly accelerated in the Minkowski vacuum. In Sec.~\ref{Final}, we summarize our conclusions. We assume metric signature $(- + + +)$ and natural units $\hbar=G=c=k_B=1$, unless stated otherwise.

\section{Interaction and decoherence}
\label{sec:Model}

Let $\phi$ be a real scalar field propagating in a globally hyperbolic spacetime ($\mathcal{M}$, $g$), where $\mathcal{M}$ is a four-dimensional manifold and $g$ is a Lorentzian metric. The action of the field is given by
\begin{equation}\label{KGaction}
    S = -\frac{1}{2}\int_{\mathcal{M}}\epsilon_\mathcal{M}(\nabla_a\phi\nabla^a\phi + m^2\phi^2 + \xi R\phi^2),
\end{equation}
where $\epsilon_\mathcal{M} \equiv \sqrt{-\det g}\text{ }dx^0 \wedge...\wedge dx^3$ is the volume 4-form, $\nabla_a$ is a torsionless covariant derivative compatible with $g$, $m$ is the field mass, and $\xi \in \mathbb{R}$ is the coupling between the field and the scalar curvature $R$. One can foliate the spacetime with Cauchy hypersurfaces $\Sigma_t$ indexed by a real parameter $t$. Thus, we define the canonical Hamiltonian of the field in any Cauchy hypersurface as
\begin{equation}
    H_\phi (t) \equiv \int_{\Sigma_t}d^3\textbf{x}(\pi(t, \textbf{x})\dot{\phi}(t, \textbf{x}) - \mathcal{L}[\phi, \nabla_a\phi]),
\end{equation}
where $\textbf{x} = (x^1, x^2, x^3)$ are spatial coordinates defined on $\Sigma_t$, $\pi(t, \textbf{x})$ is the field conjugate momentum
\begin{equation}
    \pi \equiv \frac{\delta S}{\delta \dot{\phi}},
\end{equation}
and
\begin{equation}
    \mathcal{L}[\phi, \nabla_a\phi] \equiv -\frac{\sqrt{-g\,}\;}{2}(\nabla_a\phi\nabla^a\phi + m^2\phi^2 + \xi R\phi^2),
\end{equation}
with $g\equiv \det (g_{ab})$, is the Lagrangian density. Extremizing the action in Eq.~(\ref{KGaction}) leads to the Klein-Gordon (KG) equation
\begin{equation}\label{KGeq}
    (-\nabla_a\nabla^a + m^2 + \xi R)\phi = 0.
\end{equation}

To quantize the corresponding solutions, one first defines the antisymmetric bilinear form  
\begin{equation}
    \Omega(\varphi_1, \varphi_2) \equiv \int_{\Sigma_t} \epsilon_{\Sigma}n^a(\varphi_2\nabla_a\varphi_1 - \varphi_1\nabla_a\varphi_2),
\end{equation}
on $S^{\mathbb{C}}$, the space of complex solutions of Eq.~(\ref{KGeq}). Here, $\epsilon_{\Sigma}$ is the volume 3-form on $\Sigma_t$, $n^a$ is its future-directed unit vector, and $\varphi_1, \varphi_2\in S^{\mathbb{C}}$. We then define the so-called KG~inner product as
\begin{equation}\label{KGinnerProduct}
    \langle \varphi_1, \varphi_2 \rangle \equiv -i\Omega(\overline{\varphi}_1, \varphi_2). 
\end{equation}
 This is the one adopted to build a one-particle Hilbert space, $\mathcal{H}$, from which the symmetric Fock space $\mathcal{F}_s(\mathcal{H})$ follows~\cite{W85}. The field operator is defined as

\begin{equation}\label{fieldexp}
    \phi(t, \textbf{x}) \equiv \sum_{j}\Big(u_j(t, \textbf{x})a(\overline{u}_j) + \overline{u}_j(t, \textbf{x})a^\dagger(u_j)\Big),
\end{equation}
where  $\{u_j\}$ is an orthonormal basis for $\mathcal{H}$. The operators $a(\overline{u})$ and $a^\dagger(v)$ annihilate and create modes $u\in \mathcal{H}$ and $v\in \mathcal{H}$, respectively, and satisfy
\begin{equation}\label{coeffcommut}
    [a(\overline{u}), a^\dagger (v)] = \langle u, v \rangle\mathbb{I}.
\end{equation}

The expansion in Eq.~(\ref{fieldexp}) is not mathematically well-defined. However, one can circumvent this problem by defining the field $\phi$ as an operator-valued distribution, smearing it out with a test function $f \in C^{\infty}_0(\mathcal{M})$
\begin{equation}
    \phi(f) \equiv \int_{\mathcal{M}}\epsilon_\mathcal{M}\phi(x)f(x).
\end{equation}

It can be shown that Eq.~(\ref{fieldexp}) can be cast as~\cite{W85}
\begin{equation}\label{coeffsmeared}
    \phi(f) = i\Big(a(\overline{KEf}) - a^\dagger(KEf)\Big),
\end{equation}
where
$$Ef(x) \equiv \int_{\mathcal{M}}\epsilon'_{\mathcal{M}}\Big(G^{\text{adv}}(x, x') - G^{\text{ret}}(x, x')\Big)f(x'),$$
with $G^{\text{adv}}$ and $G^{\text{ret}}$ being the advanced and retarded Green distributions associated with the operator $P\equiv(-\nabla^a\nabla_a + m^2+\xi R)$. We note that $E$ maps test functions into the space of real solutions of Eq.~(\ref{KGeq}) that have compact support when restricted to Cauchy hypersurfaces. Here, $K: S \to \mathcal{H}$ takes the positive-norm part of such solutions. It is straightforward to check that $\phi(f)$ satisfies the covariant version of the canonical commutation relations (CCR)
\begin{equation}\label{CCCR}
    [\phi(f), \phi(g)] = -i\Delta(f, g)\mathbb{I},
\end{equation}
where
\begin{equation}
    \Delta(f, g) \equiv \int_{\mathcal{M}}\epsilon_{\mathcal{M}}f(x)Eg(x).
\end{equation}
The unsmeared version is immediately obtained by noting that $\Delta(x, x') = E(x, x') = G^{\text{adv}}(x, x') - G^{\text{ret}}(x, x')$.

{We recall that the choice of the one-particle Hilbert space $\mathcal{H}$ is not unique. Still, in spacetimes with some timelike Killing field $\chi^a$, there is a ``natural" choice of $\mathcal{H}$ whose vacuum state represents the absence of particles as measured by observers following the integral curves of $\chi^a$. The no-particle state as defined by inertial observers in Minkowski spacetime is referred to as the Minkowski vacuum, $|0_M\rangle$.

Now, consider a two-level particle detector whose internal states can be represented as elements of a Hilbert space $\mathcal{H}_D$, and let the eigenvectors of the Pauli matrix $\sigma^z$, $\{| 0 \rangle, | 1\rangle\}$, be a basis of $\mathcal{H}_D$. As the gap of the detector is assumed to be degenerate, $\Delta E \to 0$, the total Hamiltonian of the system becomes $H(t) \equiv H_\phi (t) + H_{\text{int}}(t)$. The interaction term, in the interaction picture, is given by
\begin{equation}
    H_{\text{int}}^I (t) \equiv \epsilon(t)\int_{\Sigma_t}d^3\textbf{x}\sqrt{- g\,}\;\psi(t, \textbf{x})\phi(t, \textbf{x}) \otimes \sigma^z,
    \label{Hint}
\end{equation}
where $\epsilon(t) \in C^\infty_0(\mathbb{R})$ is a smooth real-valued function with compact support, modeling that the detector stays switched on for a finite-time interval, and $\psi(t, \textbf{x})$ is a smooth real-valued function, satisfying $\psi|_{\Sigma_t} \in C^{\infty}_0(\Sigma_t)$ for all Cauchy hypersurfaces $\Sigma_t$,  that models the detector extension~\cite{LM09, AL16}. We recall that the choice of $\Sigma_t$ is completely arbitrary.

We are interested now in computing the corresponding time evolution operator
\begin{equation}\label{timeevolop}
    U = T\exp\left[ -i\int_{-\infty}^{\infty}dt'H_{\text{int}}^I(t')\right].
\end{equation}
Following Ref.~\cite{AL16}, we use a Magnus expansion~\cite{BCOR09} to evaluate Eq.~(\ref{timeevolop}) non-perturbatively. Hence,
\begin{equation}
U= \exp(\Omega)
\end{equation}
such that $\Omega = \sum_{n = 1}^{\infty}\Omega_n$. The first three terms of the expansion are given by
\begin{equation}
    \Omega_1 = -i\int_{-\infty}^{\infty}dtH^I_{int}(t) = -i\phi(f)\otimes\sigma^z,
\end{equation}
\begin{equation}
    \Omega_2 = -\frac{1}{2}\int_{-\infty}^{\infty}dt\int_{-\infty}^tdt'[H^I_{int}(t), H^I_{int}(t')] = i\Xi\mathbb{I},
\end{equation}
\begin{equation}
\begin{split}
    \Omega_3 = \frac{i}{6}\int_{-\infty}^{\infty}&dt\int_{-\infty}^tdt'\int_{-\infty}^{t'}dt''\Big([H^I_{int}(t), [H^I_{int}(t'), H^I_{int}(t'')]\\
    &+ [H^I_{int}(t''), [H^I_{int}(t'), H^I_{int}(t)]\Big) = 0,
\end{split}
\end{equation}
where 
$$
f(t,{\bf x})\equiv \epsilon(t)\psi(t, {\bf x}),
$$
\begin{equation}
    \Xi \equiv \frac{1}{2}\int_{-\infty}^{\infty}dt\epsilon(t)\int_{-\infty}^{t}dt'\epsilon(t')\Delta(t, t'),
\end{equation}
and 
\begin{equation}
\begin{split}
    \Delta(t, t') \equiv \int_{\Sigma_t}d^3\textbf{x}&\sqrt{-g}\int_{\Sigma_{t'}}d^3\textbf{x}'\sqrt{-g'} \\ 
    &\times \psi(t, \textbf{x})\Delta(x, x')\psi(t', \textbf{x}'),
\end{split}
\end{equation}
and the higher-order terms can be obtained recursively. The recursive character of the Magnus expansion leads to $\Omega_n = 0$ for $n \geq 3$.  
This would not be the case if we considered the Unruh-DeWitt detector with a nonzero gap $\Delta E$. In that case, one cannot find an exact nonperturbative expression for $U$. Thus, we can write Eq.~(\ref{timeevolop}) as
\begin{equation}\label{Uoperator1sup}
    U = \exp(i\Xi)\exp(-i\phi(f)\otimes\sigma^z).
\end{equation}

As we are interested in the decoherence of spatial superpositions, let $|\gamma_1 \rangle$ and $|\gamma_2 \rangle$ $\in \mathcal{H}_{\text{path}}$ be two orthonormal (distinguishable) states centered at the classical worldlines $\gamma_1$ and $\gamma_2$, respectively, associated with the superpositions $j=1$ and $2$.  Then we extend Eq.~(\ref{Uoperator1sup}) as
\begin{equation}\label{Uoperator2sup}
    U \rightarrow \sum_{j = 1}^2U_{(j)}\otimes |\gamma_j \rangle \langle \gamma_j |,
\end{equation}
posing no dynamics on the position, where we now make
\begin{equation}
    f(t,{\bf x}) \rightarrow f_j(t,{\bf x}) \equiv \epsilon_j(t)\psi_j(t, \textbf{x}).
    \label{fj}
\end{equation}
Hence, we can define the interaction range and the coupling to the field in different ways for each component. Note that, as $\Xi$ depends on $f$, we also have to make $\Xi \rightarrow \Xi_{j}$.

Next, let us assume that the detector's state and position are entangled in the Minkowski vacuum at $t \rightarrow -\infty$ as
\begin{equation}
    | \Psi_{-\infty} \rangle = \frac{1}{\sqrt{2}}| 0_M \rangle \otimes \left(| 0 \rangle \otimes | \gamma_1 \rangle + | 1 \rangle \otimes | \gamma_2 \rangle\right),
\end{equation}
where we recall that the $|0\rangle$ and $|1\rangle$ are equal-energy states. Acting on this state with the operator $U$
in~(\ref{Uoperator2sup})   results in the asymptotic expression
\begin{equation}
    | \Psi_{\infty} \rangle =  \frac{1}{\sqrt{2}}\left[|\Phi_1 \rangle \otimes | 0 \rangle \otimes | \gamma_1 \rangle + |\Phi_2 \rangle \otimes | 1 \rangle \otimes | \gamma_2 \rangle\right],
\end{equation}
where
\begin{equation}
    | \Phi_1 \rangle = \exp(i\Xi_1)\exp(-i\phi(f_1))| 0_M \rangle,
\end{equation}
and
\begin{equation}
    | \Phi_2 \rangle = \exp(i\Xi_2)\exp(i\phi(f_2))| 0_M \rangle,
\end{equation}
with $| \Phi_1 \rangle$, $| \Phi_2 \rangle$ $\in \mathcal{F}_s(\mathcal{H}_M)$. 

Now, we follow DSW and quantify the decoherence of the detector's position through the distinguishability of the field states: 
\begin{equation}
    \mathcal{D} \equiv 1 - |\langle \Phi_1 |\Phi_2 \rangle |.
\end{equation}
The more distinguishable $| \Phi_1 \rangle$ and $| \Phi_2 \rangle$ are, the more one can infer the detector's position from measuring the scalar field, and the closer $\mathcal{D}$ approaches unity. 

By using Eq.~(\ref{CCCR}) and the Baker-Campbell-Hausdorff (BCH) identity,
\begin{eqnarray*}
  \exp(X)\exp(Y) 
  &=& 
  \exp\left(X + Y + \frac{1}{2}[X, Y] + \frac{1}{12}[X, [X, Y]] \right.
  \\
  &+& \left. \frac{1}{12}[Y, [Y, X]] + \ldots \right)  
\end{eqnarray*}
for bounded operators $X$ and $Y$, we can write 
\begin{equation}\label{fieldprod1}
|\langle \Phi_1 |\Phi_2 \rangle |=\left|\langle 0_M|\exp{[i\phi(f_1+f_2)]} |0_M \rangle\right|
\end{equation}
where the phases $\Xi_1$  and $\Xi_2$ 
do not appear due to the modulus
in \eqref{fieldprod1}.
Now, we can use Eq.~(\ref{coeffsmeared}) to expand $\phi(f_1+f_2)$ as  \begin{equation}\label{coeffsmearedsup}
    \phi(f_1+f_2) = i\Big[a(\overline{KE(f_1+f_2)}) - a^\dagger(KE(f_1+f_2))\Big],
\end{equation}
and use Eq.~(\ref{coeffcommut}) together with Zassenhaus formula,
$$
e^{X+Y}=e^{X}e^{Y}e^{-\frac{1}{2}[X,Y]},
$$
where  $[X,Y]$ is a c-number, to cast Eq.~(\ref{fieldprod1}) as 
\begin{equation}
   |\langle \Phi_1 |\Phi_2 \rangle |= \exp{\left( -\frac{1}{2}||K_M E(f_1 + f_2)||^2\right)}.
\end{equation}
Here, $K_M$ takes the positive-frequency part with respect to the inertial time. The squared norm in the right-hand side is what is called the particle number $\langle N\rangle$ of {\it entangling} (Minkowski) scalar ``photons'' defined with respect to inertial observers. Therefore
\begin{equation}
    \mathcal{D} = 1 - \exp\left({-\langle N\rangle/2}\right)
\end{equation}
with 
\begin{equation}\label{N}
    \langle N\rangle \equiv  ||K_M E(f_1 + f_2)||^2.
\end{equation}
Thus, the decoherence is caused by the emission of scalar particles that transport information away from the system. We will see that when the superposition evolves freely in Minkowski spacetime, all particle creation comes from the switching on and off of the detector. In contrast, when the superposition is uniformly accelerated, the most relevant contribution comes from the detector's interaction with the soft modes of the Unruh bath (provided it remains accelerated long enough).

To conclude this section, let us highlight an important feature of our model. If we set 
$$
\epsilon_2=\epsilon_1\equiv \epsilon\geq0,
$$ 
and take the limit $\gamma_2\rightarrow \gamma_1$, we see from Eqs.~(\ref{fj}) and~(\ref{N}) that the number of entangling scalar particles does not vanish: 
\begin{equation}
\langle N\rangle \stackrel{\gamma_2 \rightarrow \gamma_1}{=} 4\|K_M E \tilde{f}\|^2 \neq 0,
\end{equation}
where $\tilde{f}=\epsilon \psi_1$. The nonvanishing of $\langle N\rangle$ in this limit arises from the fact that, although the detector follows a well-defined trajectory, it interacts differently with the field depending on whether its initial state is $|0\rangle$ or $|1\rangle$. This can be seen directly from the interaction Hamiltonian~(\ref{Hint}) and shows that the scalar particles carry away information about the initial spin state. On the other hand, if we set
$$
\epsilon_2 = -\epsilon_1 \equiv -\epsilon,
$$ 
the number of entangling scalar particles becomes 
$\langle N\rangle = \|K_M E (\tilde{f}_1 - \tilde{f}_2)\|^2$,
where $\tilde{f}_j \equiv \epsilon \psi_j$ ($j=1,2$). Hence, in the limit $\gamma_2 \rightarrow \gamma_1$, we obtain
\begin{equation}
\langle N\rangle \stackrel{\gamma_2 \rightarrow \gamma_1}{\longrightarrow} 0.
\end{equation}
This ``erases'' the information imprinted by the initial spin on the field state and isolates the role played by the superposition of trajectories in the decoherence process, thereby establishing a direct connection to the DSW results, as we will see.  The number $ \langle N\rangle$ of entangling (Minkowski) ``scalar photons'' can be roughly regarded, in this case, as a subset of the radiated particles with wavelengths short enough (compared with the superposition separation) to permit  ``which-path identification.'' In the following, we analyze each case separately.

\section{Decoherence of inertial and uniformly accelerating detectors for
\texorpdfstring{$\epsilon_{2}=-\epsilon_{1}$}{e2 = -e1}}
\label{sec:WaldRec0}

\subsection{Inertial case}
\label{inertial-}

We start by considering the simpler case where the detector is not in superposition, in which case we make $f_2 = 0$. The coupling function in Eq.~\eqref{fj} will be defined as $\epsilon_1(t) \equiv \varepsilon \;C(t)$ with $\varepsilon = {\rm const}$ and
\begin{equation}\label{switchinginertial}
    C(t) \equiv \left\{ 
     \begin{array}{ccc} 
        e^{\alpha(t + T)}& \text{for}& t \leq -T \\
        1& \text{for}& -T \leq t \leq T \\
        e^{-\alpha(t - T)}& \text{for} &t \geq T
      \end{array}\right.,
\end{equation}
 where $1/\alpha$ is a parameter that represents how fast the detector is switched on and off, and the interaction is kept on for a proper time $2T$. We will assume that the detector is pointlike and, thus, $\psi_1(t, \textbf{x})|_{\Sigma_t} = \delta^3(\textbf{x})$. By recalling that the normalized plane waves of the one-particle Hilbert space $\mathcal{H}_M$ are
\begin{equation}
   u_{\textbf{k}} \equiv \frac{e^{i(\textbf{k}\cdot\textbf{x} - \omega_{\textbf{k}}t)}}{\sqrt{16\pi^3\omega_{\textbf{k}}}},
\end{equation}
with ${\bf k}\in \mathbb{R}^3$, $\omega_{\bf k}=\sqrt{{\bf k}^2+m^2}$, and $m$ being the field mass, we write 
\begin{equation}\label{norminer1sup}
    ||K_M Ef_1||^2 = \int d^3\textbf{k}\;  |\langle u_{\textbf{k}}, Ef_1 \rangle |^2,
\end{equation}
where we have used~\cite{LM09}
\begin{equation}
    \langle u_{\textbf{k}}, Ef_1 \rangle \equiv \frac{i\varepsilon}{\sqrt{8\pi^2\omega_{\textbf{k}}}}\tilde{C}(\omega_{\textbf{k}})
\end{equation}
with the Fourier transform of $C(t)$:
$$
\tilde{C}(\omega_{\textbf{k}}) \equiv \frac{1}{\sqrt{2\pi}}\int_{-\infty}^{\infty}dt\,  e^{i\omega_{\textbf{k}}t} \, C(t) .
$$
\begin{figure}[b]
       \centering
       \includegraphics[scale=0.4]{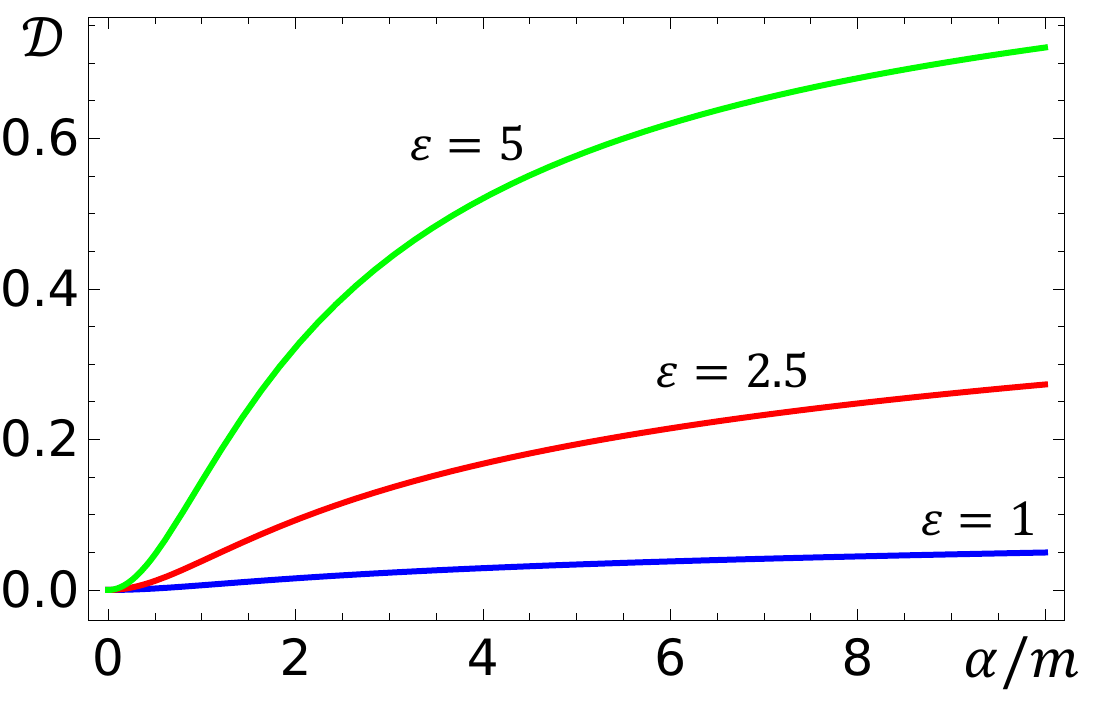}
       \caption{Decoherence as a function of the ratio $\alpha/m$ for coupling constants $\varepsilon = 1$, $\varepsilon = 2.5$, and $\varepsilon = 5$.}
       \label{Fig1_final}
\end{figure}

In the regime where the detector remains switched on for a long time relative to all other parameters, {\it i.e.}, $T \gg  1/m, 1/\alpha$, Eq.~(\ref{norminer1sup}) can be approximated by
\begin{equation}\label{deciner1sup}
    ||K_M Ef_1||^2 \approx \frac{\varepsilon^2}{2\pi^2}\left(\sqrt{1 + \left(\frac{m}{\alpha}\right)^2}\text{ arcsinh}\left(\frac{\alpha}{m}\right) - 1\right).
\end{equation}
See Fig.~\ref{Fig1_final} for the induced decoherence ${\cal D}$ as a function of $\alpha/m$. Imposing in addition $\alpha/m \gg 1$, the expression above reduces to
\begin{equation}
    ||K_M Ef_1||^2 \sim \frac{\varepsilon^2}{2\pi^2}\log \left(\frac{\alpha}{m} \right) \quad \text{for} \quad {T\gg 1/m \gg 1/\alpha}.
    \label{InertialParticleProduction1}
\end{equation}
This result is commensurate with the literature. In Ref.~\cite{HMP93}, it was shown that the corresponding emission probability of {\it massless} scalar particles in the Minkowski vacuum from an inertial detector with energy gap $\Delta E > 0$ is
\begin{equation}
    {\cal P}^M \sim \frac{\varepsilon^2}{2\pi^2}\log \left(\frac{\alpha}{\Delta E} \right) \quad \text{for} \quad T\gg 1/\Delta E \gg 1/\alpha.
\end{equation}
(Note that $I_{\rm in}$ and $I_{\rm abs}$ in Eq.~(19) of Ref.~\cite{HMP93} vanish for inertial detectors ($a\to 0$), and one can use Eq.~(27) of that paper to write $I_{\rm sp}$ for $T\gg 1/\Delta E \gg 1/\alpha$.) This means that an infinite amount of energy is necessary to instantaneously ($1/\alpha \to 0$) switch a detector with a finite energy gap ($\Delta E > 0$), accounting for the creation of an infinite number of particles. The same formula indicates that, regardless of whether the detector is continuously ($1/\alpha > 0$) switched, a divergent production of {\it massless} particles is expected for gapless detectors ($\Delta E \to 0$). Equation~\eqref{InertialParticleProduction1} conveys the same message by beginning with a gapless detector ($\Delta E = 0$) and taking $m\to 0$ in the end. The physical interpretation is that it turns out to be arbitrarily easy to excite a gapless detector coupled to massless scalar particles. This explains why we consider, from the outset, massive scalar fields in this paper.

Now, we consider the case where the detector is in a superposition of inertial paths far from each other, separated by a distance $L={\rm const}$---see Fig.~\ref{Fig2_final}. In this case, 
\begin{equation}\label{psi1psi2inertial}
\psi_1(t, \textbf{x})|_{\Sigma_t} = \delta^3(\textbf{x}), 
\quad
\psi_2(t, \textbf{x})|_{\Sigma_t} \equiv \delta^3(\textbf{x} -L\,\hat{\textbf{x}}),
\end{equation}
with $\hat{\textbf{x}} \equiv (1,0,0)$ and  $\epsilon_1(t) = -\epsilon_2(t)$,  where $\epsilon_1 \equiv \varepsilon \;C(t)$ and $C(t)$ is given by Eq.~(\ref{switchinginertial}). For the large~$T$ regime, $T \gg 1/\alpha,\, 1/m,\, L$, a similar procedure results in 
\begin{eqnarray}\label{deciner2sup-}
    && ||K_M E (f_1 + f_2)||^2 
    = 
    \frac{\varepsilon^2 \alpha^2}{\pi^2 m^2}
    \nonumber \\
   &\times& 
    \int^{\infty}_0 d\tilde{k}\frac{\tilde{k}^2\left[ 1 - \text{sinc}\left(mL\sqrt{\tilde{k}^2 + 1}\right)\right]}{\left( \tilde{k}^2 + 1 \right)^{3/2}\left(\tilde{k}^2 + 1 + (\alpha/m)^2\right)},
\end{eqnarray} 
where $\text{sinc}(x) \equiv \sin(x)/x$. Let us now go a step further and consider the regime $T \gg 1/\alpha\gg 1/m\gg L$ of Eq.~~\eqref{deciner2sup-}, where 
\begin{equation}\label{inertialapprox-}
    ||K_M E(f_1 + f_2)||^2 \sim \varepsilon^2{\alpha^2L^2}.
\end{equation}
Figure~\ref{Fig3_final} depicts Eq.~\eqref{deciner2sup-}. We see a plateau for $L \gg 1/\alpha$. This is because the wavelength of the emitted entangling Minkowski particles scales with the switching time $1/\alpha$. Thus, for $L \gg 1/\alpha$, the superposition paths interfere little with one another. On the other hand, for $L\ll 1/\alpha$, the smaller $L$ is, the more the (destructive) interference contributes to suppressing $\langle N \rangle$ [see Eq.~\eqref{inertialapprox-}]. (Recall that $\epsilon_2=-\epsilon_1$, here.) This is expected, since for small enough $L$, it becomes very hard to distinguish the path that sources the radiation, thereby suppressing decoherence.
\begin{figure}
       \centering
       \includegraphics[scale=0.5]{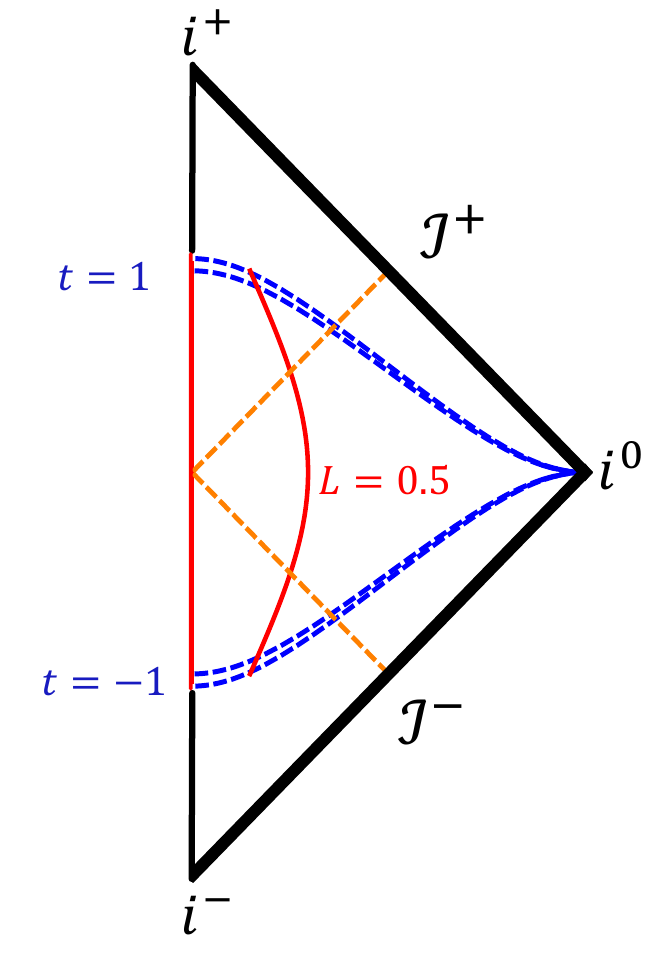}
       \caption{The worldlines represent the detector's inertial superposition that are $L=0.5$ far away from each other in Minkowski spacetime. The blue dashed lines represent constant proper-time hypersurfaces for a congruence of inertial observers.  The detector remains switched on for a proper time interval $T=2$, and the switching process lasts on the order of $1/\alpha = 0.1$.
       }
       \label{Fig2_final}
\end{figure}
\begin{figure}[b]
       \centering
       \includegraphics[scale=0.45]{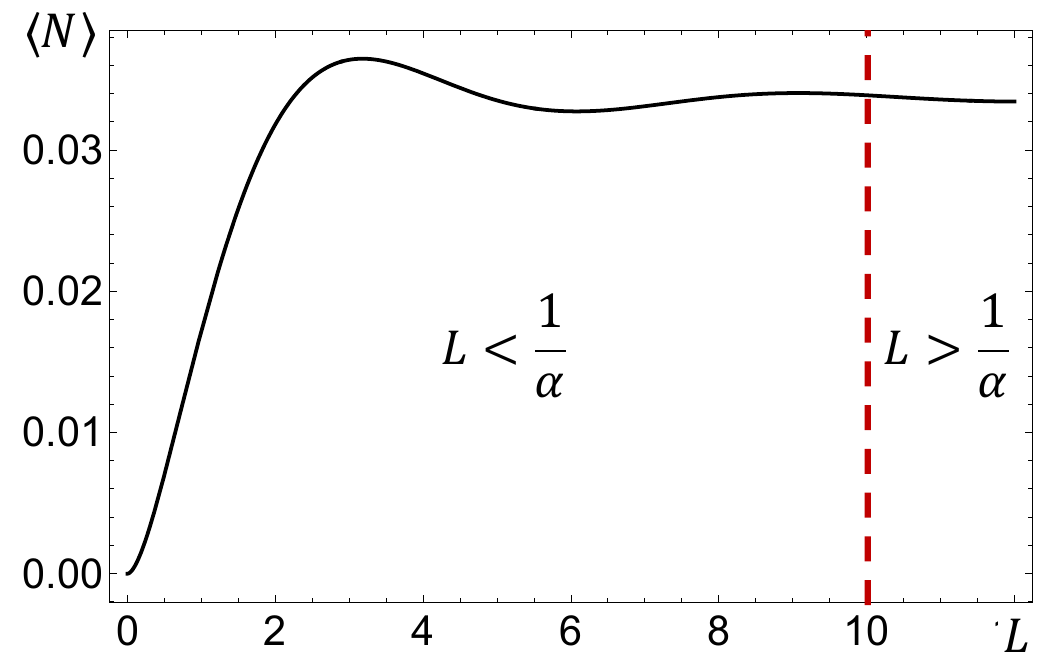}
       \caption{Expected number of entangling Minkowski particles caused by the detector's switching in terms of the distance~$L$ between the spatial components for $\varepsilon = 10$, $1/\alpha = 10$, and $m = 1$. We see that for $L\ll 1/\alpha$, the superposed paths interfere destructively in the emission of entangling particles. On the other hand, in the region $L \gg 1/\alpha$, the paths are sufficiently separated so that interference is suppressed. } 
       \label{Fig3_final}
\end{figure}

Equation~\eqref{inertialapprox-} is similar to the one obtained in Ref.~\cite{DSW23} for superposed inertial charges, where the emission of entangling particles, 
\begin{equation}\label{waldN-}
\langle N \rangle \sim q^2 L^2/[\text{min}(T_1, T_2)]^2,
\end{equation}
occurs due to the forward and reverse Stern-Gerlach stages that are presupposed to happen with a proper acceleration $A$. Comparing Eq.~(\ref{waldN-}) with Eq.~(\ref{inertialapprox-}), we observe that $\varepsilon$ plays the role of the charge $q$, informing how strong the corresponding interaction is, and $1/\alpha$ corresponds to  $\text{min}(T_1, T_2)$, reflecting the action of the external agent responsible for the decoherence, and setting the time scale for which the phenomenon becomes relevant.  In both situations, one can minimize the decoherence by making the process adiabatic. In the detector case, this means taking $\alpha$ small compared to $m$. However, we will see that this does not prevent soft-radiation emission when one analyzes the same scenario in a spacetime that contains Killing horizons, such as Rindler space. 

\subsection{Uniformly accelerated case}
\label{acc-}

Now, let us investigate the detector's decoherence when the components of the superposition are uniformly accelerated in Minkowski spacetime with $\epsilon_2=-\epsilon_1$. It is suitable to analyze the problem in the right Rindler wedge~(RRW) covered with Rindler coordinates ($\tau$, $\xi$, $y$, $z$),  $\tau, \xi, y, z \in \mathbb{R}$~\cite{CHM08,W84}:
\begin{equation}
        t = \frac{e^{a\xi}}{a}\sinh(a\tau), \quad
        x = \frac{e^{a\xi}}{a} \cosh(a\tau).
\end{equation}
See the highlighted region depicted in Fig. \ref{Fig4_final} defined by $x>|t|$. In these new coordinates, the line element is given by
\begin{equation}\label{Rds2}
    ds^2 = e^{2a\xi}(-d\tau^2 + d\xi^2) + dy^2 + dz^2.
\end{equation}
We can similarly define the so-called left Rindler wedge (LRW) as the region $x<-|t|$. Like the RRW, the LRW can be covered by Rindler coordinates, in which the metric takes the form given in Eq.~(\ref{Rds2}).
\begin{figure}[t]
       \centering
       \includegraphics[scale=0.5]{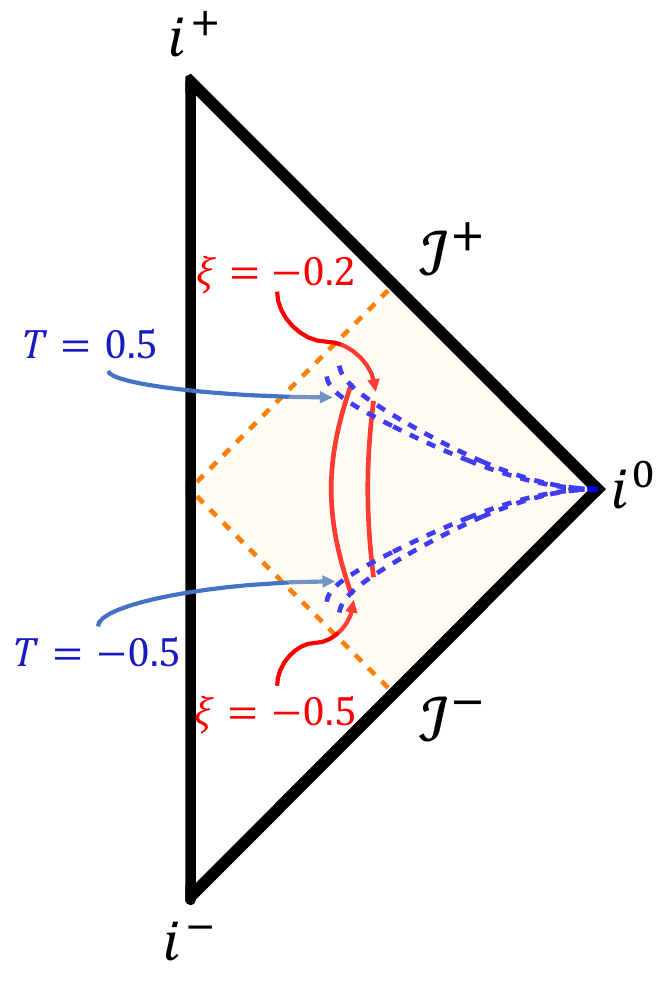}
       \caption{The worldlines represent the detector's uniformly accelerated superposition in the right-Rindler wedge at $\xi =-0.2$ and $\xi=-0.5$ (${\bf x}_\perp={\rm const}$). The blue dashed lines represent constant-proper-time hypersurfaces for the congruence of Rindler observers.  The detector remains switched on for a proper time interval of $2T=1$, and the process of switching on and off lasts on the order of $1/\alpha = 0.1$. We are interested in studying how decoherence varies as the superposition distance changes in $\xi$.}
       \label{Fig4_final}
\end{figure}

It can be seen from Eq.~(\ref{Rds2}) that the time translations $\varphi_{\hat{\tau}}(\tau, \xi, y,z)\equiv (\tau+\hat{\tau}, \xi, y,z)$ define an isometry and, thus, {$\zeta^a=(\partial_\tau)^a$} is a time-like Killing vector field in the RRW. As a result, we can define an orthonormal basis $\{v^R_{\omega\textbf{k}_\perp}\}$ of solutions of the KG equation~\eqref{KGeq} which vanish in the LRW and have positive frequency with respect to the Killing time $\tau$ in the RRW as given by
\begin{equation}\label{rightrindlermodes}
    v^R_{\omega\textbf{k}_\perp} \equiv \left[\frac{\sinh(\pi\omega/a)}{4\pi^4a}\right]^{1/2} K_{i\omega/a}\left(\frac{e^{a\xi}}{a} \sqrt{k^2_\perp + m^2\,} \right) e^{i\textbf{k}_\perp \cdot \textbf{x}_\perp} e^{-i\omega\tau},
\end{equation}
where $\textbf{x}_\perp \equiv (y, z)$, $\textbf{k}_\perp \in \mathbb{R}^2$, $\omega > 0$, and $K_{\nu}(x)$ is the modified Bessel function of the second kind. With the above right-Rindler modes, one can define the left-Rindler modes by the relation
\begin{equation}
    v^L_{\omega\textbf{k}_\perp}(t, x, \textbf{x}_\perp) \equiv \overline{v^R_{\omega\textbf{k}_\perp}}(-t, -x, \textbf{x}_\perp),
\end{equation}
which, in turn, have positive frequency with respect to the Killing time $\tau$ in the LRW and vanish on the RRW. Note that, in terms of the Rindler coordinates that cover the LRW, these modes take the form given in Eq.~(\ref{rightrindlermodes}). The modes  $v^R_{\omega\textbf{k}_\perp}$ and $ v^L_{\omega\textbf{k}_\perp}$, together with their complex conjugates, form a complete set of solutions of the KG equation~\eqref{KGeq} for the entire Minkowski spacetime.    

It is noteworthy that the modes $v^R_{\omega\textbf{k}_\perp}$ and $ v^L_{\omega\textbf{k}_\perp}$ can be combined to construct a complete set of modes that are purely positive-frequency modes with respect to the inertial time $t$, thereby forming a complete basis for the one-particle Hilbert space naturally associated with inertial observers. This construction is achieved through the so-called Unruh modes
\begin{equation}\label{UnruhModes}
        w^1_{\omega\textbf{k}_\perp} \equiv \dfrac{v^R_{\omega\textbf{k}_\perp} + e^{-\pi\omega/a}\overline{v^L_{\omega-\textbf{k}_\perp}}}{\sqrt{1 - e^{-2\pi\omega/a}}}, \quad
        w^2_{\omega\textbf{k}_\perp} \equiv \dfrac{v^L_{\omega\textbf{k}_\perp} + e^{-\pi\omega/a}\overline{v^R_{\omega-\textbf{k}_\perp}}}{\sqrt{1 - e^{-2\pi\omega/a}}},
\end{equation}
satisfying $w^1_{-\omega\textbf{k}_\perp} = w^2_{\omega\textbf{k}_\perp}$. They are particularly interesting for our purposes because, although they have positive frequency with respect to inertial time $t$, they are labeled by quantum numbers associated with observables measured by accelerated observers. As a result, they turn out to be quite helpful in connecting the physical descriptions of inertial and accelerated observers.

Before we proceed to compute the number of emitted entangling Minkowski scalar particles, we need to define the switching function that will be used in this case.
 Analogously to the inertial case~(\ref{switchinginertial}), we will choose
\begin{equation}\label{switchingaccelerated}
    C(\tau) \equiv  \left\{ 
     \begin{array}{ccc} 
        e^{\alpha[\tau \exp(a\xi) + T]} &  \text{for} & \tau e^{a\xi}\leq -T \\
        1 &\text{for}    &-T \leq \tau e^{a\xi} \leq T \\
        e^{-\alpha[\tau \exp(a\xi) - T]} &\text{for}& \tau e^{a\xi} \geq T
    \end{array},
    \right.
\end{equation}
ensuring that the detector remains switched on for an interval ${2}T$ and the switching processes last about $1/\alpha$ in proper time, irrespective of the worldline $\xi=\text{const}$.  

Analogously to Eq~\eqref{psi1psi2inertial} of the inertial case, we now write
\begin{equation}\label{psi1}
    \psi_1(\tau,\xi,{\bf x}_\perp)|_{\Sigma_\tau} \equiv \frac{1}{\sqrt{- g}}\delta(\xi)\delta(y)\delta(z)
    \end{equation}
and
\begin{equation}\label{psi2}
\psi_2(\tau,\xi,{\bf x}_\perp)|_{\Sigma_\tau} \equiv \frac{{e^{a\xi}}}{\sqrt{- g}}\delta(\xi - \tilde{\xi})\delta(y)\delta(z),
\end{equation}
where $\Sigma_{\tau}$ are $\tau=\text{const}$ Cauchy hypersurfaces for the RRW. 

Now, by using the Unruh modes~(\ref{UnruhModes}), we write
\begin{equation}
\begin{split}
    &||K_M E(f_1 + f_2)||^2 = \int_0^\infty d\omega\int_{\mathbb{R}^2} d^2\textbf{k}_\perp \\
    &\Big[|\langle w^1_{\omega\textbf{k}_\perp}, E(f_1 + f_2)\rangle |^2 + |\langle w^2_{\omega\textbf{k}_\perp}, E(f_1 + f_2)\rangle |^2\Big],
\end{split}
\end{equation}
and using the relation between the Unruh modes $w^2_{\omega \textbf{k}_\perp} = w^1_{-\omega\textbf{k}_\perp}$, we obtain
\begin{equation}
    ||K_M E(f_1 + f_2)||^2 = \int_{\mathbb{R}}d\omega\int_{\mathbb{R}^2} d^2\textbf{k}_\perp|\langle w^1_{\omega\textbf{k}_\perp}, E(f_1 + f_2)\rangle |^2,
\end{equation}
where
\begin{eqnarray}
\!\!\!\!\!   \langle w^1_{\omega\textbf{k}_\perp}, E(f_1 + f_2)\rangle  
    & =& \frac{i\varepsilon \exp[{\pi \omega/(2a)}]}{\sqrt{4\pi^3 a\,}}
    \nonumber \\ 
    &\times&  \left[K_{-i\omega/a}\Big(\sqrt{k^2_\perp + m^2}\Big/a\Big)\tilde{C}_1(\omega) \right.
    \nonumber \\
     &\textcolor{blue}{-}& \left. {e^{a\tilde{\xi}}}K_{-i\omega/a}\Big(e^{a\tilde{\xi}} \sqrt{k^2_\perp + m^2} \Big/ a\Big)\tilde{C}_2(\omega)\right]\nonumber\\
\end{eqnarray}
with $\tilde{C}_1(\omega)$ and $\tilde{C}_2(\omega)$ being the Fourier transform of $C_1(\tau)$ and $C_2(\tau)$, respectively, {\it i.e.},
$$\tilde{C}_i(\omega) \equiv \frac{1}{\sqrt{2\pi}}\int_{-\infty}^{\infty}d\tau\,  e^{i\omega \tau} \, C_i(\tau), \quad i=1,2.
$$
Therefore, one ends up with
\begin{equation}\label{KEf1-f2}
    ||K_M E(f_1 + f_2)||^2 = \frac{\varepsilon^2a}{\pi^2}(I_1 + I_2 - I_{\text{int}}), 
\end{equation}
with
\begin{equation} \label{I1}
    I_1 \equiv \int^{\infty}_{0}d\omega \cosh(\pi\omega/a)|\tilde{C}_1(\omega)|^2\int_{m/a}^{\infty}dx x|K_{i\omega/a}(x)|^2,
\end{equation}
\begin{equation} \label{I2}
    I_2 \equiv  \int_0^\infty d\omega \cosh(\pi\omega/a)|\tilde{C}_2(\omega)|^2  
   \int_{me^{a\xi}/a}^\infty dx x|K_{i\omega/a}(x)|^2,
\end{equation}
 and
\begin{eqnarray} \label{Iint}
        I_{\text{int}} &\equiv& 2 {e^{a\tilde{\xi}}}\int_0^{\infty}d\omega \cosh(\pi\omega/a)\tilde{C}_1(\omega)\tilde{C}_2(\omega) \nonumber \\
        &\times& \int_{m/a}^{\infty }dx\; x \;\text{Re} \left[K_{i\omega/a}\Big(x \Big)K_{-i\omega/a}\left( x e^{a\tilde{\xi}}\right) \right],
\end{eqnarray}
where $I_1$ and $I_2$ are the contributions coming from the $\xi = 0$ and $\xi = \tilde{\xi}={\rm const}$ worldlines, respectively, and  $I_{\text{int}}$ stems from the interference between them.

We are particularly interested in comparing our results with those of DSW, who assume that the components evolve for a proper time much larger than any other scales of the problem, {\it i.e.}, $T \gg 1/a, d$, where $a$ stands here for the proper acceleration of the superposition paths. (Here, they disregard the acceleration~$A$ over the forward and reverse Stern-Gerlach stages.) Note that because DSW also assumes that the proper distance between the paths, $d=(e^{a\tilde{\xi}}-1)/a$, is much smaller than the distance between them and the horizon, $1/a$, they are in the regime where $1/a \gg d \approx \tilde{\xi}$, and both components have approximately the same proper acceleration. In this regime, DSW find that the number of emitted entangling photons (associated with the emission/absorption of soft photons to/from the Hawking thermal bath outside the event horizon) goes as~\cite{DSW23} 
\begin{equation}
   \langle N \rangle \sim q^2\, a^3\, d^2\, T.
   \label{DSW}
\end{equation}
\begin{figure}[t]
       \centering
       \includegraphics[scale=0.4]{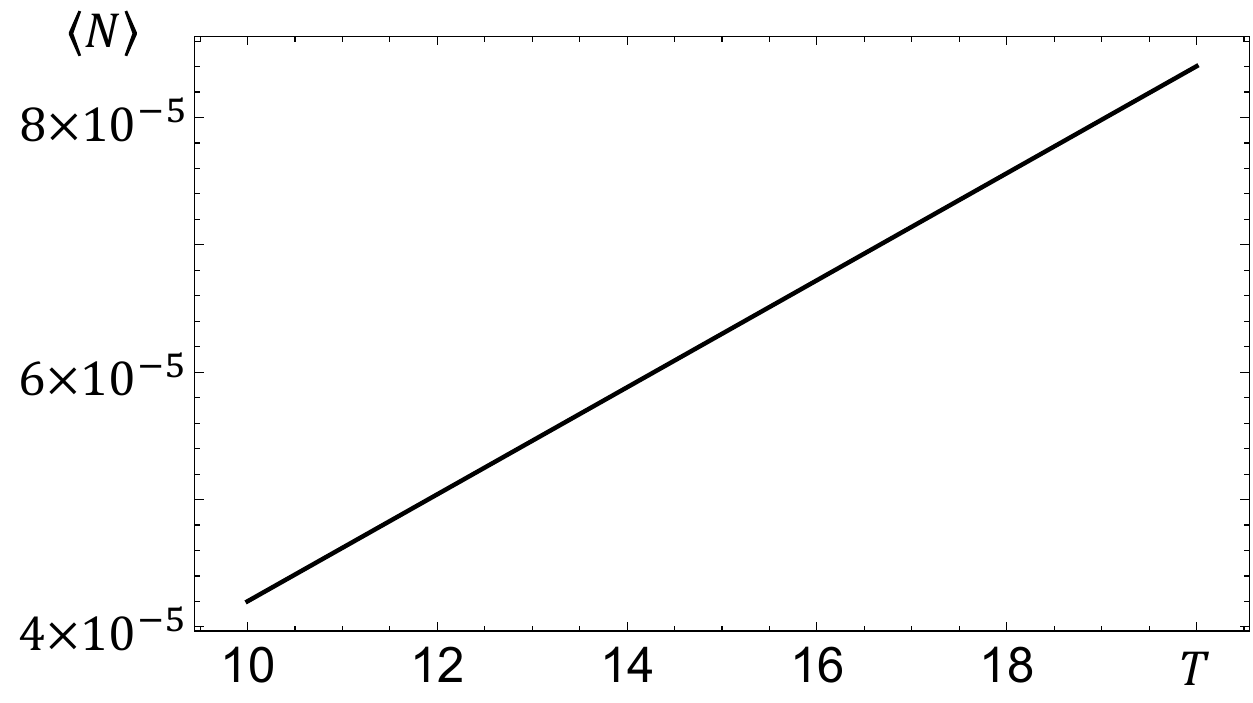}
       \caption{Expectation number of emitted Minkowski entangling particles as a function of ~$T$. The graph assumes $\varepsilon = 1$, $m = 0$, $\alpha=10^{-5}$, $a = 1$, and $\tilde{\xi} = 0.01$, satisfying $1/m \gg 1/\alpha \gg T \gg 1/a \gg d$. The line grows linearly as expected [see Eq.~\eqref{DSW}].}
       \label{Fig5_final}
\end{figure}

Analogously, we will focus on the regime $T \gg 1/a \gg d$ to enable comparison with their results. Furthermore, we switch the detector on and off smoothly and reduce the field mass to a scale smaller than any other in the problem, leading us to the comparison regime    
$$
1/m \gg 1/\alpha \gg T \gg 1/a \gg d.
$$
In Figs.~\ref{Fig5_final},~\ref{Fig6_final}, and~\ref{Fig7_final}, we plot the expectation number of emitted entangling Minkowski particles 
\begin{equation} \label{<N>}
\langle N \rangle \equiv ||K_M E(f_1 + f_2)||^2  -||K_M E(f_1 + f_2)||^2\Big|_{T\to 0},
\end{equation}
as defined by inertial observers over the time interval that the detector remains fully switched on. The subtraction in Eq.~\eqref{<N>} roughly discounts the excess of entangling photons due to emission during the detector switching on/off processes. This simulates the DSW regime, where the emission of entangling photons during the Stern-Gerlach processes is negligible. (No subtraction is introduced in the inertial case because the whole decoherence comes from the switching of the detector. Similarly, in the DSW inertial case, all decoherence comes from the Stern-Gerlach processes. Accordingly, discounting the emission of entangling photons over the switching of the detector or the Stern-Gerlach processes in the inertial case would erase the decoherence.) This allows us to largely relax the constraints on $\alpha$. Moreover, for the accelerated case, we can plot the graphs assuming $m\to 0$ from the outset, thereby recovering the DSW's regime of interest: 
$$
T \gg 1/a \gg d.
$$
We recall that the radiation emitted by uniformly accelerated sources, as defined by inertial observers, corresponds in the Rindler frame to the absorption and emission of Rindler particles with arbitrarily small frequencies~\cite{higuchi_1992R, higuchi_1992, portales_oliva_2022, vacalis_2023, portales_oliva_2024, brito_2024}; this is why the DSW decoherence may be attributed to the source's response to soft particles of the Hawking thermal bath.  

First, we see from Fig.~\ref{Fig5_final} that~$\langle N \rangle$ [see Eq.~\eqref{<N>}] grows linearly with the total time $2T$ over which the detector is fully switched on, in agreement with Eq.~\eqref{DSW}. Also, we have checked that each path $\gamma_1$ and $\gamma_2$ contributes to the particle emission as 
\begin{equation}
\varepsilon^2 \frac{2T\Delta E}{2\pi} \left( \frac{1}{e^{2\pi \Delta E/a_0} -1} +\lambda \right) \stackrel{\Delta E\to 0} {\longrightarrow} 
\varepsilon^2 \frac{ 2T a_0}{4 \pi^2} .
\label{scalarsource}
\end{equation}
The left-hand side of Eq.~\eqref{scalarsource} gives the ``excitation'' ($\lambda=0$) and ``deexcitation'' ($\lambda=1$) probabilities for an Unruh-DeWitt detector with energy gap $\Delta E$ and proper acceleration $a_0$, where the switching of the detector is carried out adiabatically. The excitation and deexcitation of the detector are accompanied by the absorption and emission of a Rindler particle with $\omega=\Delta E$, respectively, both corresponding to the emission of a Minkowski particle as viewed by inertial observers~\cite{CHM08}. In our case, the detector is gapless, so we must take the limit $\Delta E \to 0$. Thus the right-hand side of Eq.~\eqref{scalarsource} is the expected number of massless scalar Minkowski particles emitted by the gapless detector in the inertial vacuum~\cite{LFM19}. 
\begin{figure}[t]
       \centering
       \includegraphics[scale=0.4]{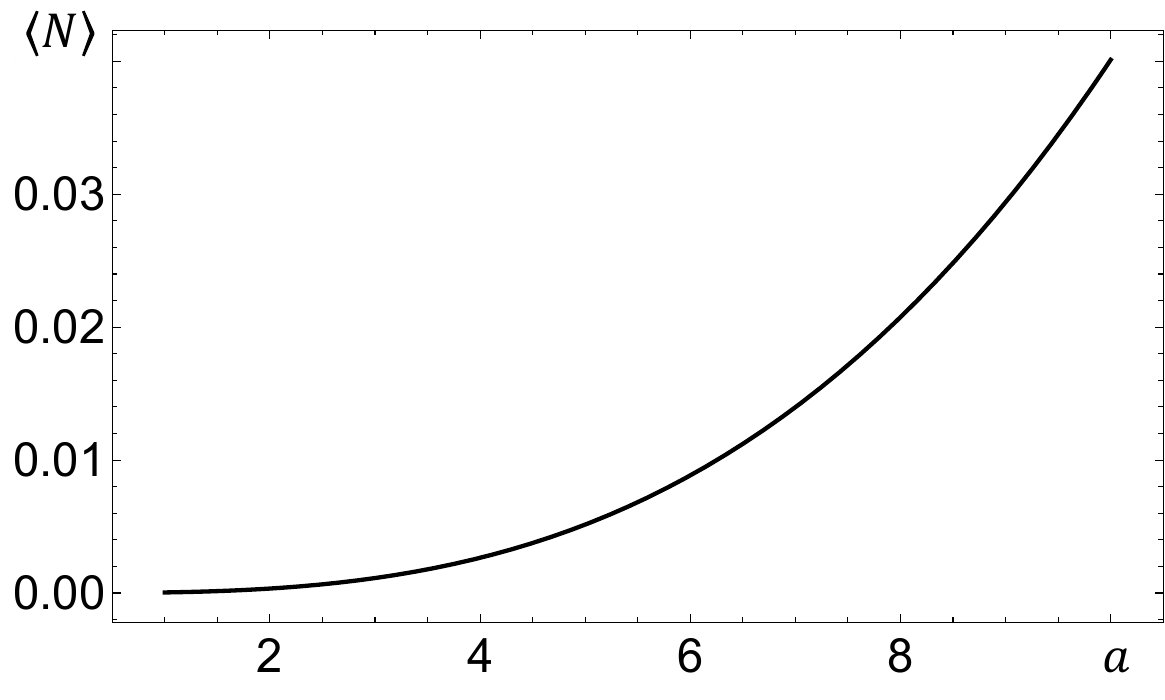}
       \caption{ Expectation number of Minkowski particles as a function of $a$ (proper acceleration for $\xi=0$). The graph assumes $\varepsilon =1$, $m = 0$, $\alpha = 10^{-5}$, $T = 10$, and $\tilde{\xi} = 0.01$. The curve fits Eq.~\eqref{DSW} with proportionality constant $8.1\times 10^{-2}$.}
       \label{Fig6_final}
\end{figure}

In Fig.~\ref{Fig6_final} we plot~$\langle N \rangle$ as a function of $a$. It is important to note that as we augment the proper acceleration $a$ of the path at $\xi=0$, we are also augmenting the proper distance $d= a^{-1}(e^{a\tilde{\xi}}-1)$ between the superpositions, but we can disregard it as far as $a\tilde{\xi} \ll 1$, since in this case $d\approx \tilde{\xi}$. Again, each path contributes to the particle emission as given by Eq.~\eqref{scalarsource}. We emphasize that the curve obeys $T \gg 1/a \gg d$ and fits DSW's Eq.~\eqref{DSW} (differing from each other by less than $1\%$ along the curve for a proportionality constant $8.1\times 10^{-2}$).
\begin{figure}[th]
       \centering
       \includegraphics[scale=0.4]{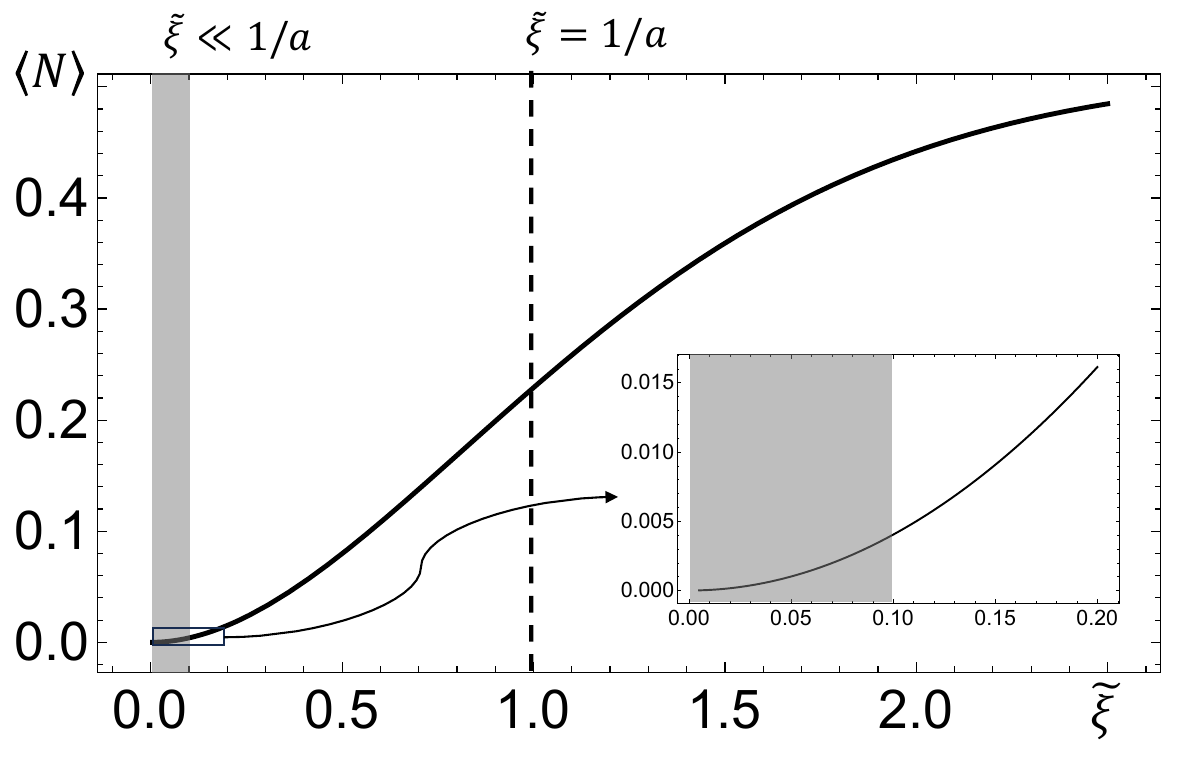}
       \caption{Expectation number of emitted entangling Minkowski particles (as defined by inertial observers) as a function of $\tilde{\xi}$ (which corresponds to the proper distance $d$ between the paths for $\tilde{\xi}\ll 1/a$). The graph assumes $\varepsilon =1$, $m = 0$, $\alpha = 10^{-5}$, $T = 10$, and $a = 1$. The curve fits Eq.~\eqref{DSW} with proportionality constant $8.1\times 10^{-2}$ in the DSW's regime $1/a \gg \xi \approx d$ (see gray region in the inset).}
       \label{Fig7_final}
\end{figure}

In Fig.~\ref{Fig7_final} we plot~$\langle N \rangle$ as a function of $\tilde{\xi}$. We recall that for $\tilde{\xi} \ll 1/a$, $\tilde{\xi}$ plays the role of the proper distance $d$ between the superposition paths. Otherwise, $d= a^{-1}(e^{a\tilde{\xi}}-1)$. For a large enough proper distance, $\tilde{\xi} \gg 1/a$,  the distance  $d$ becomes much larger than the typical wavelength of the emitted {Minkowski} particles, suppressing the (destructive) interference and leading to large decoherence. On the other hand, for $\tilde{\xi} \ll 1/a$, complying with DSW's inequality $T \gg 1/a \gg d$ (see gray region in the graph), interference takes over, and the curve {fits DSW's Eq.~\eqref{DSW} (differing from each other by less than $1\%$ along the curve for a proportionality constant $8.1 \times 10^{-2}$)}.

\section{Decoherence of inertial and uniformly accelerating detectors for
\texorpdfstring{$\epsilon_{2}=\epsilon_{1}$}{e2 = e1}}
\label{sec:WaldRec+}

For the sake of completeness, let us analyze the analogous case of Sec.~\ref{sec:WaldRec0} for $\epsilon_2= \epsilon_1$, and show that it differs from the DSW results.

\subsection{Inertial case}

First, we analyze the case in which $\epsilon_2 = \epsilon_1$, with 
 $\epsilon_1(t) \equiv \varepsilon \, C(t)$, $\varepsilon = {\rm const}$, and $C(t)$ given by Eq.~(\ref{switchinginertial}). As in the previous case, we begin by considering a detector prepared in a superposition of inertial trajectories separated by a fixed distance $L = {\rm const}$---see Fig.~\ref{Fig2_final}. As in Eq.~\eqref{psi1psi2inertial}, we have
\begin{equation*}\label{psi2psi1inertial}
\psi_1(t, \textbf{x})|_{\Sigma_t} = \delta^3(\textbf{x}), 
\quad
\psi_2(t, \textbf{x})|_{\Sigma_t} \equiv \delta^3(\textbf{x} -L\,\hat{\textbf{x}}),
\end{equation*}
with $\hat{\textbf{x}} \equiv (1,0,0)$. { Then, instead of Eq.~\eqref{deciner2sup-}, we end up here with}
\begin{eqnarray}\label{deciner2sup}
    && ||K_M E (f_1 + f_2)||^2 
    = 
    \frac{\varepsilon^2 \alpha^2}{\pi^2 m^2}
    \nonumber \\
   &\times& 
    \int^{\infty}_0 d\tilde{k}\frac{\tilde{k}^2\left[ 1 + \text{sinc}\left(mL\sqrt{\tilde{k}^2 + 1}\right)\right]}{\left( \tilde{k}^2 + 1 \right)^{3/2}\left(\tilde{k}^2 + 1 + (\alpha/m)^2\right)},
\end{eqnarray} 
for the large~$T$ regime, $T \gg 1/\alpha,\, 1/m,\, L$. {Going further and assuming $T \gg 1/\alpha\gg 1/m\gg L$ in Eq.~\eqref{deciner2sup}, we get}
\begin{equation}\label{inertialapprox+}
    ||K_M E(f_1 + f_2)||^2 \sim {\varepsilon^2\alpha^2}/{m^2} .
\end{equation}
The smaller the mass $m$, the easier it is to create particles, and the larger the decoherence. In Fig.~\ref{Fig8_final}, we plot this for $\varepsilon=10$, $1/\alpha=10$, $1/m=1$ as a function of $L$. We see that in the region $L \gg 1/\alpha$, the superposition paths are sufficiently separated so that one path does not influence the other during the switching process, which explains the plateau {stabilization at the same value as in Fig.~\ref{Fig3_final}}. This is in contrast to the regime $L\ll 1/\alpha$, where each path influences the other during the switching process. We emphasize that the difference with Fig.~\ref{Fig3_final} occurs because, although the superposition degenerates in the limit $L \to 0$, the scalar field still radiates away information about the particle's initial spin state when $\epsilon_2 = \epsilon_1$, thereby contributing to decoherence (recall discussion at the end of Sec.~\ref{sec:Model}).  
\begin{figure}[th]
       \centering
       \includegraphics[scale=0.45]{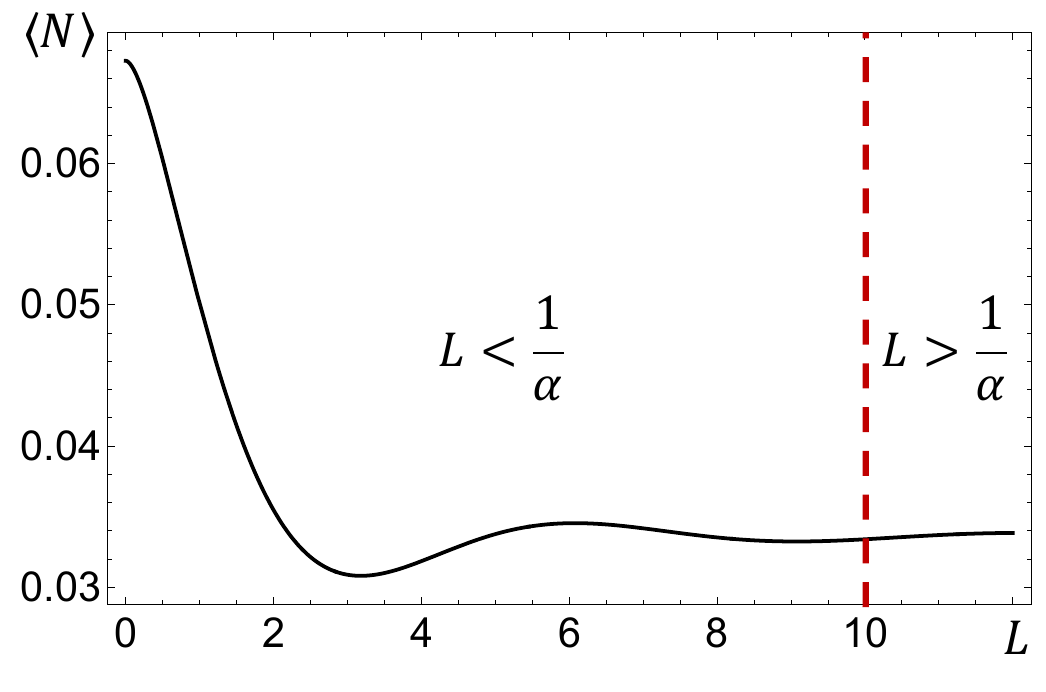}
       \caption{Expected number of entangling particles (Eq.~\eqref{deciner2sup}) caused by the detector's switching in terms of the distance between the spatial components~$L$ for $\varepsilon = 10$, $1/\alpha = 10$, and $1/m = 1$. We see that for $L\ll 1/\alpha$, the superposed paths interfere constructively in the emission of entangling Minkowski particles. On the other hand, in the region $L \gg 1/\alpha$, the paths are sufficiently separated so that the interference is suppressed. In this region, 
        the values of $\langle N \rangle$ in
       Figs.~\ref{Fig3_final} and~\ref{Fig8_final} approach each other.}
       \label{Fig8_final}
\end{figure}

\subsection{Uniformly accelerated case}

Now, let us investigate the detector's decoherence when the components of the superposition are uniformly accelerated in Minkowski spacetime. As in Sec.~\ref{acc-}, it is suitable to analyze the problem in the right Rindler wedge~(RRW) covered with Rindler coordinates ($\tau$, $\xi$, $y$, $z$),  $\tau, \xi, y, z \in \mathbb{R}$. By using Eqs.~(\ref{switchingaccelerated})-(\ref{psi2}) and following the same steps that led to Eq.~(\ref{KEf1-f2}) but recalling that $\epsilon_2=\epsilon_1$ here, we write
\begin{equation}
    ||K_M E(f_1 + f_2)||^2 = \frac{\varepsilon^2a}{\pi^2}(I_1 + I_2 + I_{\text{int}}), 
\end{equation}
where $I_1$, $I_2$, and $I_{\rm int}$ are given in Eqs.~\eqref{I1}, \eqref{I2}, and \eqref{Iint}, respectively. In Figs.~\ref{Fig9_final},~\ref{Fig10_final}, and~\ref{Fig11_final}, we plot the expectation number of emitted entangling Minkowski particles 
$$
\langle N \rangle \equiv ||K_M E(f_1 + f_2)||^2  -||K_M E(f_1 + f_2)||^2\Big|_{T\to 0},
$$
as defined by inertial observers over the time interval that the detector remains fully switched on, and compare them with Figs.~\ref{Fig5_final}, \ref{Fig6_final}, and~\ref{Fig7_final} of Sec.~\ref{acc-}.

We see from Fig.~\ref{Fig9_final} that~$\langle N \rangle$ grows proportionally to the interaction time $2T$, as expected. Each path contributes to the particle emission as given by Eq.~\eqref{scalarsource}, and half of the particle emission in Fig.~\ref{Fig9_final} comes from the interference term. In contrast to Fig.~\ref{Fig3_final}, the particle emission in Fig.~\ref{Fig9_final} is much more vigorous because the interference term contributes constructively.
\begin{figure}[th]
       \centering
       \includegraphics[scale=0.42]{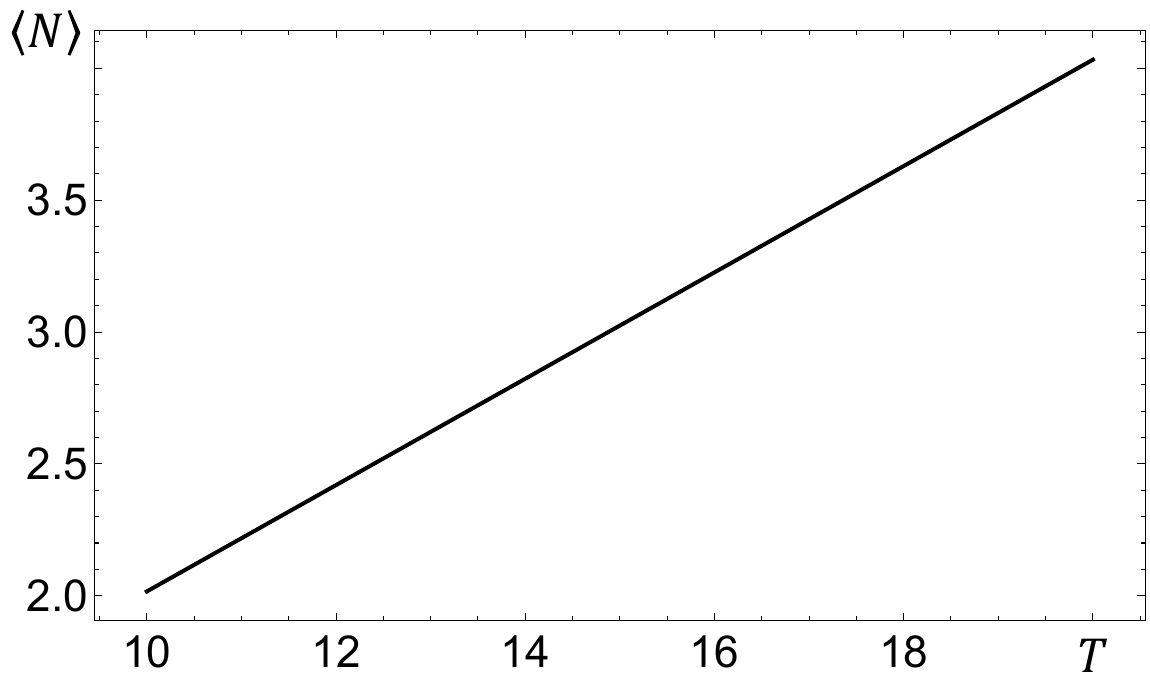}
       \caption{Expectation number of entangling Minkowski particles as a function of ~$T$. The graph assumes $\varepsilon = 1$, $m = 0$, $\alpha=10^{-5}$, $a = 1$, and $\tilde{\xi} = 0.01$. The line grows linearly as expected.}
       \label{Fig9_final}
\end{figure}

In Fig.~\ref{Fig10_final} we plot~$\langle N \rangle$ as a function of $a$. (The parameters chosen satisfy $a\tilde{\xi} \ll 1$, keeping the proper distance between the superposition paths  constant irrespective of the variation of the proper acceleration $a$ of the path at $\xi=0$.) Again, each path contributes to the particle emission as given by Eq.~\eqref{scalarsource}, and half of the particle emission comes from the interference term. From the perspective of Rindler observers, decoherence increases because the superposition interacts with an Unruh bath at a higher temperature $a/2\pi$, which has more soft modes. Note how the interference term changes the overall behavior of Fig.~\ref{Fig10_final} in comparison to Fig.~\ref{Fig6_final}. Figure~\ref{Fig10_final} is in line with Ref.~\cite{LMW26}, which concludes at the first order of perturbation that the decoherence rate scales linearly with the acceleration for a case corresponding to $\epsilon_1 = \epsilon_2$.
\begin{figure}[th]
       \centering
       \includegraphics[scale=0.42]{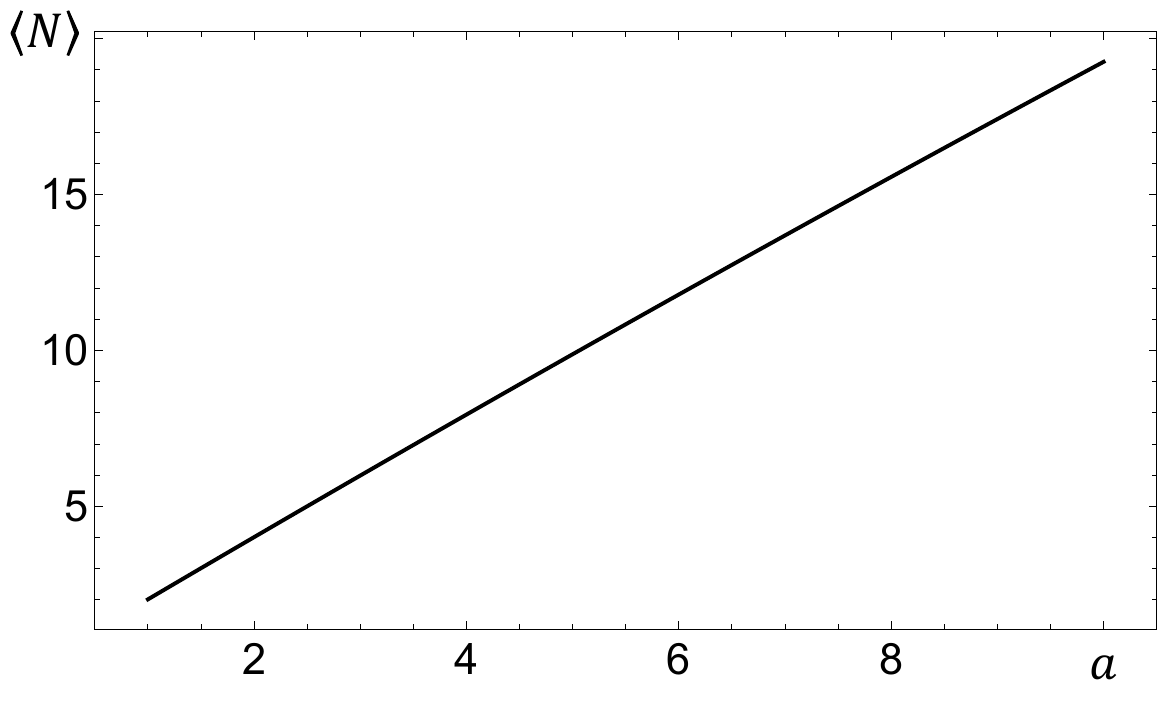}
       \caption{Expectation number of entangling Minkowski particles as a function of the proper acceleration $a$. The graph assumes $\varepsilon =1$, $m = 0$, $\alpha = 10^{-5}$, $T = 10$, and $\tilde{\xi} = 0.01$. }
       \label{Fig10_final}
\end{figure}

Finally, Fig.~\ref{Fig11_final} shows that the detector's decoherence decreases as the proper distance between the paths increases. In contrast to the previous case (see Fig.~\ref{Fig7_final}), maximum constructive interference occurs when the paths coincide, and it gradually diminishes as the proper distance between them grows. For $\tilde{\xi}\gg 1/a$, the curve goes asymptotically to a value corresponding to the emission of entangling Minkowski particles only due to the path at $\xi=0$. (Recall that for $\tilde{\xi}\gg 1/a$, the interference term is damped, and the path at $\tilde{\xi}$ has a proper acceleration too small to interact with soft Rindler particles of the Unruh thermal bath, which concentrate at the horizon.)
\begin{figure}[th]
       \centering
       \includegraphics[scale=0.42]{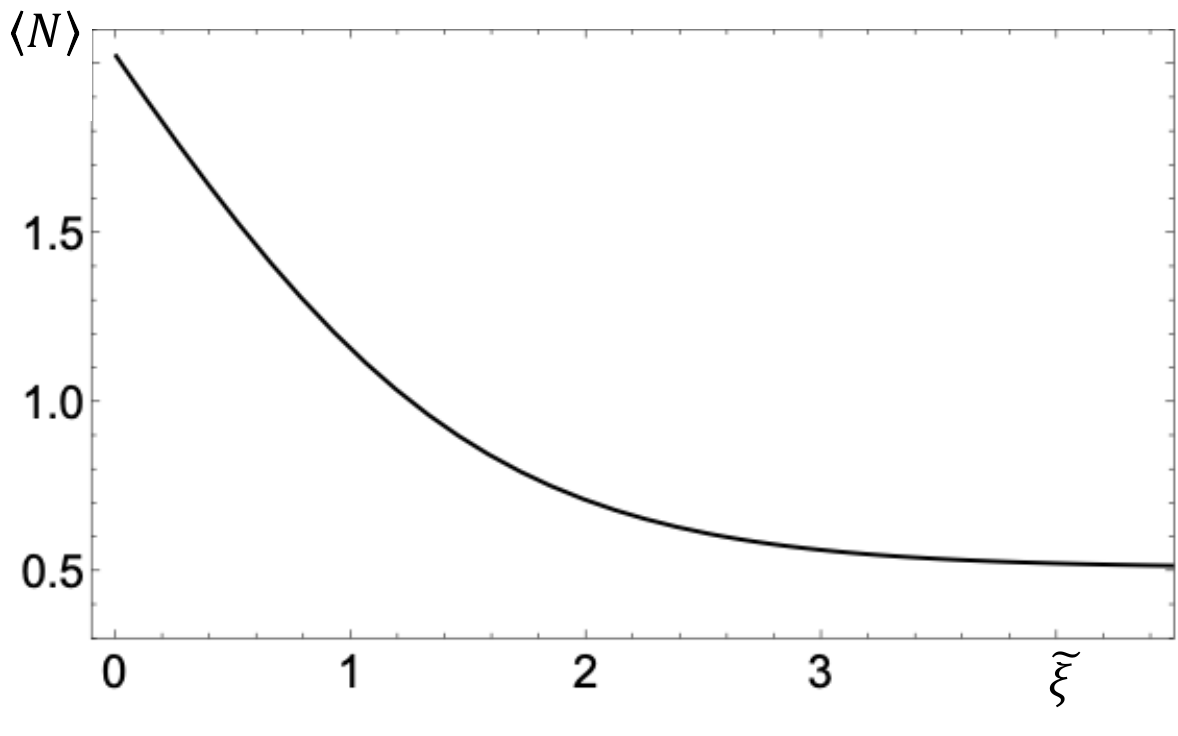}
       \caption{Expected number of entangling Minkowski particles as a function of $\tilde{\xi}$. The graph assumes $\varepsilon =1$, $m = 0$, $\alpha = 10^{-5}$, $T = 10$, and $a = 1$. The larger the $\tilde{\xi}$, the smaller the contributions from (i)~the interference term and (ii)~the path at $\tilde{\xi}$, leading to a decrease of $\langle N \rangle$ up to a nonzero minimum value. }
       \label{Fig11_final}
\end{figure}

   \section{Discussion and closing remarks }
   \label{Final}

Inspired by DSW's papers~\cite{DSW22a, DSW22b, DSW23, DSW25}, we have analyzed the decoherence of a gapless Unruh-DeWitt detector superposed in uniformly accelerated paths in Minkowski spacetime. The detector is assumed to interact with a massive quantum scalar field. The field is initially set to the Minkowski vacuum, and the forward and reverse Stern-Gerlach processes used to split and recombine the charge in DSW were replaced by switching the detector on and off. Since the combined evolution of the detector and field system is exact in our case, our model can be used to analyze several aspects of the DSW mechanism in a more controlled manner.

In our setup, decoherence is caused by the emission of entangling Minkowski scalar particles from the system, carrying information about both the path superposition and the initial spin state. As a result, in order to effectively ``isolate''  the role played by the path superposition on the decoherence process, we have first analyzed the case where the coupling constants $\epsilon_1$ and $\epsilon_2$ associated with the superposition paths following $\gamma_1$ and $\gamma_2$, respectively, satisfy $\epsilon_2=-\epsilon_1$. Our Figs.~\ref{Fig5_final}, \ref{Fig6_final}, and~\ref{Fig7_final} are in good agreement with DSW's decoherence formula~\eqref{DSW} in the proper region 
$$
T \gg 1/a \gg d,
$$
but goes beyond it (see, {\it e.g.}, Fig.~\ref{Fig7_final} for $ \tilde{\xi} \nll 1/a$, corresponding to $1/a \ngg d$). The DSW decoherence can be traced back to the fact that radiation emitted by uniformly accelerated sources, as measured by inertial observers, corresponds in the Rindler frame to the absorption and emission of Rindler particles with arbitrarily small frequencies ({\it i.e.}, zero-energy Rindler particles)~\cite{higuchi_1992R, higuchi_1992, portales_oliva_2022, vacalis_2023, portales_oliva_2024, brito_2024}.  

We also analyzed the case where $\epsilon_2=\epsilon_1$, in which case the information about the path and about the initial state gets intertwined in the scalar radiation. As in the previous case, we have shown that $\langle N\rangle$  is proportional to the interaction proper time $T$ when  $T\gg 1/a, d$. We have also studied how $\langle N\rangle$ varies with the proper acceleration $a$, making clear the difference between the cases $\epsilon_2=\epsilon_1$ and $\epsilon_2=-\epsilon_1$ (compare Figs.~\ref{Fig6_final} and~\ref{Fig10_final}). Finally, as we analyze the behavior of $\langle N\rangle$ as a function of the proper distance $d$, we found that we have complete constructive interference when $d\rightarrow 0$, making $\langle N\rangle$ attain its maximum value, which gradually decreases as $d$ increases. This reflects the damping of the interference term and the ceasing of emission from the displaced path at $\tilde{\xi}$ as $d \rightarrow \infty$, as expected.

\acknowledgments

L.B.N.B. is supported by EPSRC Grant EP/W524694/1 and STFC Grant ST/Y509644/1, and was supported by Coordenação de Aperfeiçoamento de Pessoal de Nível Superior (CAPES) Grant 88887.947554/2024-00 and the Perimeter Scholars International Scholarship 2024/25. The research of R.B.M. is supported by the Natural Sciences and Engineering Research Council of Canada. G.E.A.M. was partially supported by the National Council for Scientific and Technological Development (CNPq) and S\~ao Paulo Research Foundation (FAPESP) under grants~301508/2022-4 and~2022/10561-9, respectively. 

\section*{DATA AVAILABILITY}

{The data that support the findings of this article are openly available~\cite{data}.}

\bibliography{references}

\end{document}